\documentclass[lettersize,journal]{IEEEtran}

\usepackage{amsmath,amsfonts}
\usepackage{algorithmic}
\usepackage{algorithm}
\usepackage{array}
\usepackage{xcolor}
\usepackage[caption=false,font=normalsize,labelfont=scriptsize,textfont=scriptsize]{subfig}
\usepackage{textcomp}
\usepackage{url}
\usepackage{bbding}
\usepackage{verbatim}
\usepackage{stfloats}
\usepackage{graphicx}
\usepackage{cite}
\usepackage{booktabs}
\usepackage {enumitem}
\usepackage{makecell}
\newcounter{magicrownumbers}
\newcommand\rownumber{\stepcounter{magicrownumbers}\arabic{magicrownumbers}}
\begin{document}

\title{K-Means Based Constellation Optimization for Index Modulated Reconfigurable Intelligent Surfaces}

\author{Hao Liu, Jiancheng An, \IEEEmembership{Member,~IEEE,} Wangyang Xu, Xing Jia, Lu Gan, and Chau Yuen, \IEEEmembership{Fellow,~IEEE}
        % <-this % stops a space
\thanks{This research is supported by the Ministry of Education, Singapore, under its MOE Tier 2 (Award number MOE-T2EP50220-0019). Any opinions, findings and conclusions or recommendations expressed in this material are those of the author(s) and do not reflect the views of the Ministry of Education, Singapore. This work is partially supported by Sichuan Science and Technology Program under Grant 2023YFSY0008 and 2023YFG0291.}
\thanks{H. Liu, X. Jia, W. Xu and L. Gan are with the School of Information and Communication Engineering, University of Electronic Science and Technology of China (UESTC), Chengdu, Sichuan 611731, China. H. Liu, X. Jia and L. Gan are also with the Yibin Institute of UESTC, Yibin, Sichuan 644000, China (E-mail: liu.hao@std.uestc.edu.cn, xingjia1999@163.com, wangyangxu@std.uestc.edu.cn, ganlu@uestc.edu.cn). J. An is with the Engineering Product Development Pillar, Singapore University of Technology and Design, Singapore 487372 (e-mail: jiancheng\_an@sutd.edu.sg). C. Yuen is with the School of Electrical and Electronics Engineering, Nanyang Technological University, Singapore 639798 (e-mail: chau.yuen@ntu.edu.sg).}\vspace{-0.8cm}
}
\maketitle

\begin{abstract}
Reconfigurable intelligent surface (RIS) has recently emerged as a promising technology enabling next-generation wireless networks. In this letter, we develop an improved index modulation (IM) scheme by utilizing RIS to convey information. Specifically, we study an RIS-aided multiple-input single-output (MISO) system, in which the information bits are conveyed by reflection patterns of RIS rather than the conventional amplitude-phase constellation. Furthermore, the K-means algorithm is employed to optimize the reflection constellation to improve the error performance. Also, we propose a generalized Gray coding method for mapping information bits to an appropriate reflection constellation and analytically evaluate the error performance of the proposed scheme by deriving a tight upper bound of the average bit error rate (BER). Finally, numerical results verify the accuracy of our theoretical analysis as well as the substantially improved BER performance of the proposed RIS-based IM transmission scheme.
\end{abstract}


\section{Introduction}
\IEEEPARstart{R}{econfigurable} intelligent surface (RIS) has recently attracted significant interest as a promising technology enabling next-generation wireless networks \cite{6g}. Specifically, an RIS is a programmable metasurface composed of a large number of passive elements, each of which can independently reconfigure the amplitude and/or phase of the incident signals cost-efficiently\cite{ref3}. As such, RIS is capable of significantly enhancing the signal strength and mitigating the interference with less hardware cost and energy consumption, thus improving the spectrum and energy efficiency of the communication systems \cite{An1}. Benefiting from these appealing advantages, RIS is considered a key technology to enable future energy-efficient communication networks\cite{ref10}.

Reflection optimization constitutes a big challenge to unlock the full potential of RIS. In order to address this challenge, the authors of \cite{ref3} proposed an alternating optimization (AO) method to jointly design the downlink precoding and reflection optimization for minimizing the transmit power of the RIS-assisted multiuser multiple-input single-output (MISO) systems. In \cite{shu2022beamforming}, the authors derived an approximate expression of secrecy rate and proposed three methods to jointly optimize the phase shifts of RIS and transmit power at transmitter to maximize the secrecy rate in RIS-aided secure spatial modulation (SM) system. In \cite{ref20}, the block coordinate descent method was employed to jointly design the transmit beamformer and RIS reflection coefficients of a wideband multiuser MISO system. Considering the excessive pilot overhead for acquiring a huge number of reflected channel coefficients, the authors of \cite{An1} proposed a codebook-based transmission protocol capable of striking flexible trade-offs between performance, pilot overhead, and implementation complexity of an RIS-aided multiuser system.

Besides enhancing the channel quality and mitigating the interference, RIS can also be used as a carrier to convey information \cite{An2,ref11, zhang2022irs, ref13, quadrature, 9956998, 9610069}. 
For example, the authors of \cite{ref11} proposed a novel reflecting modulation scheme by simultaneously employing the transmit signals and RIS reflection patterns to convey information. 
In \cite{zhang2022irs}, the authors proposed a generalized space shift keying in RIS-aided MIMO system as well as a pre-greedy aided maximum likelihood detector.
Following this, the authors of \cite{ref13} studied an RIS-aided quadrature reflection modulation by partitioning the RIS elements into two subsets and reflecting the incident signals in two orthogonal directions.
The authors of \cite{9956998} used index modulation (IM) in a non-orthogonal multiple access (NOMA) system to improve the spectrum efficiency.
Furthermore, the authors of \cite{9610069} combined the RIS and spatial scattering modulation (SSM) in a millimeter-wave uplink system to attain low bit error rate (BER).
Besides, in \cite{9684431}, the authors analyzed the path loss of RIS-aided SM with an ON/OFF pattern. An interpolation-based signature constellation was proposed in \cite{hussein2021reconfigurable}.

Nonetheless, the aforementioned research efforts generally focus on increasing the system's spectrum efficiency or enhancing the received SNR. 
The constellation optimization for improving the BER performance of the RIS-aided IM scheme remains largely unexplored.
Against this background, we develop a novel IM scheme with improved BER performance for the RIS-assisted downlink MISO system by optimizing the reflection constellation. Specifically, first of all, the K-means algorithm is employed to divide all legitimate reflection patterns into several clusters. Based on the clustering results, we propose an effective scheme to select the appropriate reflection patterns for increasing the sum of distances between all potential constellation pairs. Following this, we develop a generalized Gray coding method for mapping the information bits to the corresponding reflection constellation. In addition, we provide an approximate expression to characterize the average BER performance of the proposed RIS-aided IM scheme. Finally, numerical results verify our theoretical analysis as well as the improved BER performance of the proposed RIS-aided IM scheme compared to existing transmission schemes without performing any constellation optimization.

\emph{Notation}: vectors, matrices and sets are denoted by bold-face lower-case, bold-face upper-case and upper-case calligraphic letters, respectively. $\mathbf{A}^H$, $\mathbf{A}^T$ and $||\mathbf{A}||_F$ denote Hermitian transpose, transpose and Frobenius norm of $\mathbf{A}$, respectively. $|\cdot|$ represents the modulus. $\text{diag}(\boldsymbol{\xi})$ denotes a diagonal matrix with the entries of $\boldsymbol{\xi}$ on its main diagonal. $\mathbb{C}^{x \times y}$ denotes the space of $x \times y$ complex-valued matrices. $n \sim \mathcal{CN}(0, \rho^2)$ represents a circularly symmetric complex Gaussian (CSCG) random scalar with zero mean and $\rho^2$ variance. $P(a \to b)$ represents the pairwise error probability (PEP) detecting $a$ to $b$. $Q(\cdot)$ denotes the Q-function.

\vspace{-1mm}
\section{System Model}
\begin{figure}[!t]
\centering
\includegraphics[width=0.7\linewidth]{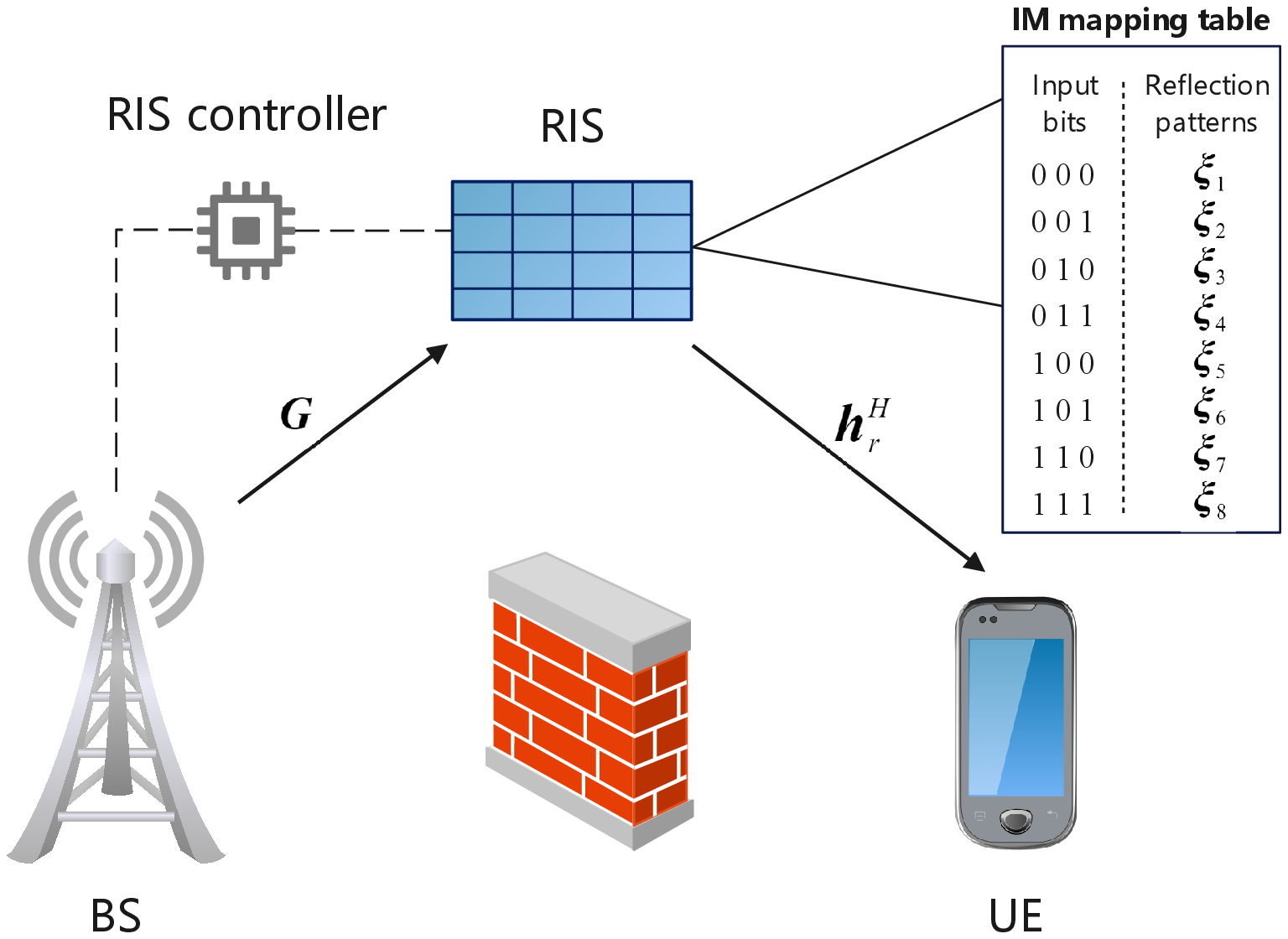}
\caption{Schematic of an RIS-assisted downlink MISO system where index modulation with $L=8$ is considered.}
\label{fig1}
\end{figure}

As shown in Fig. \ref{fig1}, we consider an RIS-aided downlink MISO system, where an RIS having $N$ reflecting elements is deployed to enhance the communications from a multiple-antenna base station (BS) to a single-antenna user equipment (UE). The number of transmit antennas at the BS is denoted by $M$. Furthermore, each RIS element can reflect the incident signals with an independent phase shift by tuning a smart RIS controller\cite{6g}. Let $ \boldsymbol{\xi}= [\xi_1, \xi_2,...,\xi_N]^T \in \mathbb{C}^{N\times 1}$ denote the reflection pattern \cite{ris_pattern} vector of the RIS, where $ \xi_n=e^{j\phi_n} $ is the reflection coefficient of the $n$-th RIS element, with $\phi_n$ denoting the corresponding phase shift and $j^2=-1$. In addition, let $ \boldsymbol{\Xi}=\text{diag}(\boldsymbol{\xi}) \in \mathbb{C}^{N\times N} $ denote the corresponding reflection pattern matrix. 
In this letter, we consider that each RIS element can only impose discrete phase shift\cite{An1}. Let $\mathcal{B}=\{0, p, 2p, ..., (B-1) p\} $ denote the feasible phase shift set, where $ p=2\pi/B $ and $B$ is the number of discrete phase shift levels. Thus we have $R=B^N$ total reflection patterns and $ \phi_n\in \mathcal{B}$. 

Moreover, let $ \textbf{G}\in \mathbb{C}^{N\times M}$ and $\textbf{h}_r^H \in \mathbb{C}^{1\times N} $ denote the channel from the BS to the RIS and that from the RIS to the UE, respectively. Let $x$ denote the normalized signal transmitted from the BS, satisfying $|x|^2=1$. As a result, the end-to-end composite channel from the BS to the UE can be expressed as $ \mathbf{h}^H=\mathbf{h}_r^H\mathbf{\Xi G}=\boldsymbol{\xi}^T\text{diag}(\mathbf{h}_r^H)\mathbf{G}=\boldsymbol{\xi}^T \mathbf{Z} \in \mathbb{C}^{1\times M}$, where $\mathbf{Z}=\text{diag}(\mathbf{h}^H_r) \mathbf{G} \in \mathbb{C}^{N \times M}$ denotes the cascaded reflected channel. Thus the received signal at the UE can be expressed as
\begin{equation}
\label{equ1}
y=\sqrt{\rho} \mathbf{h}^H \mathbf{w} x+n = \sqrt{\rho} \boldsymbol{\xi}^T \mathbf{Z} \mathbf{w} x+n,
\end{equation}
where  $\mathbf{w} \in \mathbb{C}^{M \times 1}$ denotes the transmit beamforming vector at the BS, $\rho$ denotes the average transmit power at the BS and $n \sim \mathcal{CN}(0, \sigma^2)$ denotes the additive white Gaussian noise (AWGN) at the UE. In this letter, we adopt the maximum ratio transmission (MRT) to design the transmit beamforming vector \cite{ref3}, i.e., $\mathbf{w}=\mathbf{h} / \lVert \mathbf{h} \rVert$.

Instead of existing schemes enhancing the communication quality by designing an appropriate RIS reflection pattern, in this letter, we consider an RIS-aided IM scheme by conveying information via the reflection patterns. Specifically, let $\boldsymbol{\xi}_1, \boldsymbol{\xi}_2, ..., \boldsymbol{\xi}_L $ denote the selected reflection patterns where $L=2^b$ is the number of reflection patterns selected for conveying $b$ information bits. The specific method for selecting $L$ reflection patterns will be detailed in Section III. 

At the receiver, in order to pursue the optimal performance of the proposed scheme, we adopt the optimal maximum likelihood (ML) demodulation, which is expressed as
\begin{equation}
\label{equ4}
\hat{l}_{ML} =\text{arg}\underset{l}{\text{min}} \; 
||y-\sqrt{\rho} \boldsymbol{\xi}_l^T \mathbf{Z} \mathbf{w}_l x_l ||^2_F,
\end{equation}
where $\hat{l}_{ML}$ denotes the reflection pattern index detected by the ML detector, while $\mathbf{w}_l$ and $x_l$ denote the transmit beamforming vector and transmit signal associated with the $l$-th reflection pattern $\boldsymbol{\xi}_l$, respectively. Note that designing appropriate transmit signals $x_l$ associated with different reflection patterns might further increase the average distance of all constellation pairs, which will be further discussed in Section III.

\section{The Proposed Constellation Optimization Method}
In this section, we will consider the constellation optimization by selecting $L$ appropriate reflection patterns for conveying information from total $R$ candidates at the transmitter.
In contrast to existing IM schemes selecting reflection patterns without any optimization \cite{ref11, hussein2021reconfigurable, 9956998}, we endeavour to increase the average distance sum of constellation to improve the error performance. Specifically, we first design the transmit signal for increasing the average distance sum between all candidate constellation points. Then the classic K-means clustering technique is employed to divide all reflection patterns available into $L$ clusters, based on which we select the appropriate reflection constellation. Moreover, we propose a generalized Gray coding method for mapping the information bits to the corresponding reflection constellation.
\vspace{-0.4cm}
\subsection{Reflection Constellation Design}
\subsubsection{Generation of a Universal Composite Channel Set}

Let $g_r = \boldsymbol{\xi}_r^T \mathbf{Zw}_r x_r \in \mathbb{C},\, r=1,2,...,R,$ denote the effective symbol and $\mathcal{G}=\{g_1, g_2, ..., g_R\}$ denote the effective symbol set. Note that for the sake of maximizing the received power, the beamforming vector $\mathbf{w}_r$ endeavors to align all $\boldsymbol{\xi}_r^T\mathbf{Z}$. Hence, in order to further improve the error performance of the proposed RIS-aided IM scheme, one needs to design appropriate transmit signals $x_r$ to maximize the Euclidean distance between the effective symbols. In this letter, for the sake of brevity, we adopt a heuristic solution by applying the straightforward symmetric constraint. Specifically, we first sort $g_r$'s in descending order of the channel gain, i.e., $|g_r|$. Let $g'_r$ denote the sorted version of $g_r$. Due to the symmetry of the reflection pattern set $\mathcal{B}^N$, the adjacent sorted symbols taking the inverses RIS reflection pattern have the identical channel gain. Hence, we consider the symmetric constraint by rotating $x_r$ in the odd position of $g_r'$ with a random phase shift while rotating the succeeding symbol with an inverse phase shift. More specifically, we have 
\begin{equation}
\label{rotation}
\left\{  
             \begin{array}{ll}
            x_r=e^{j\theta}, \,\theta \sim \mathcal{U}[0,\pi),&\text{if} \;\, r= 1,3,...,R-1,\\
			x_r = x_{r-1}e ^{j\pi},&\text{if} \;\, r=2,4,...,R,
             \end{array}  
\right .
\end{equation}
respectively, where $\mathcal{U}$ denotes the uniform distribution and $x_r$ is the transmit signal associated with $g_r'$. Thus, we apply the symmetric constraint to obtain $R$ effective symbols.

\subsubsection{The K-means Clustering}
Next, we divide the $R$ effective symbols into $L$ clusters by applying the K-means clustering technique. Specifically, we randomly choose the first centroid $c_1$ among the symbol set $\mathcal{G}$. For each $g_r \in \mathcal{G}, r=1,2,...,R,$ we evaluate the minimum distance between $g_r$ and all selected centroids, i.e., 
\begin{equation}
\label{assign}
D_r = \underset{i}{\text{min}} ||g_r-c_i|| ,\, i =1,2, ...,l-1.
\end{equation}
Following this, we select the $g_{\tilde r}$ having maximal $D_r$ as the $l$-th centroid $c_l$, i.e.,
\begin{equation}
\label{next_center}
\tilde r = \text{arg}\underset{r}{\text{max}}\, D_r.
\end{equation}
Repeating (\ref{assign}) and (\ref{next_center}) for $L-1$ times until generating other $L-1$ centroids.
 
After initializing $L$ centroids, we then assign constellation to the nearest centroid, i.e.,
\begin{equation}
\label{assign_centroid}
\mathcal{C}_l^{(t)} = \{ g_r: ||g_r - c_l^{(t)}||^2 \leq ||g_r - c_i^{(t)}||^2 , 1 \leq i \leq L \},
\end{equation}
where $\mathcal{C}_l^{(t)}$ and $c_l^{(t)}$ denote the $l$-th cluster set and the $l$-th centroid, respectively, of the $t$-th iteration. Following this, we update each centroid by averaging the received symbols in each cluster as follows
 \begin{equation}
 \label{new_centroid}
 c_l^{(t+1)} = \frac{1}{|\mathcal{C}_l^{(t)}|} \underset{g_r\in \mathcal{C}_l^{(t)} }{\sum} g_r.
 \end{equation}
After repeating (\ref{assign_centroid}) and (\ref{new_centroid}) for several times, the centroids gradually tend to convergence, and we obtain $L$ clusters with clustering center vector $\mathbf{c}$.

\begin{algorithm}[tb]
\caption{Reflection Constellation Optimization Based on K-means Clustering.}
\label{kmeans}
\begin{algorithmic}
\STATE 
\STATE \rownumber. \textbf{Input}: $\boldsymbol{\xi}_1, \boldsymbol{\xi}_2, ..., \boldsymbol{\xi}_R, \mathbf{Z}$.
\STATE \rownumber. \hspace{0.5cm}  Generate $R$ effective symbols $g_1, g_2, ..., g_R$ by (\ref{rotation}).
\STATE \rownumber. \hspace{0.5cm} Initialize $L$ centroids by applying (\ref{assign}) and (\ref{next_center}).
\STATE \rownumber. \hspace{0.5cm} Update the centroids by applying (\ref{assign_centroid}) and (\ref{new_centroid}) until \\ \hspace{0.9cm} the convergence is achieved.
\STATE \rownumber. \hspace{0.5cm}   Select the appropriate reflection pattern by applying \\ \hspace{0.9cm} (\ref{equ6}) and (\ref{equ7}).
\STATE \rownumber. \hspace{0.5cm} Sort the selected reflection pattern by applying (\ref{gray_coding}).
\STATE \rownumber. \hspace{0.5cm} Map sequence $\hat{\mathbf{g}}$ to $b$-bit Gray codes.
\STATE \rownumber. \textbf{Output}: $\hat{\mathbf{g}}$.
\end{algorithmic}
\label{alg1}
\end{algorithm}

\subsubsection{Selecting Appropriate Reflection Constellation}
It is natural to note that using the cluster centroids directly may not maximize the sum distance of all constellation pairs. Hence, according to the clustering results, we need perform additional procedures to select appropriate reflection constellation points. Specifically, we first select a cluster set $\mathcal{C}_l$ randomly and calculate the sum distance between the effective symbol $g_k \in \mathcal{C}_l, k=1,2,...,|\mathcal{C}_l|$ and all other cluster centroids in $\mathbf{c}$. Specifically, the sum distance of the $k$-th effective symbol in the $l$-th cluster is given by
\begin{equation}
\label{equ6}
d_{k,l} = \underset{c_i \in \mathbf{c},c_i \neq c_l}{\sum} d(g_k, c_i),
\end{equation}
where $d(g_k, c_i)$ denotes the Euclidean distance between the symbol $g_k$ and the clustering center $c_i$. Thus the $l$-th constellation is determined by selecting the one having the maximal value of $d_{k,l}$, i.e.,
\begin{equation}
\label{equ7}
\tilde{k}_l = \text{arg}\underset{k}{\text{max}}\,d_{k,l},\, k=1,2,...,|\mathcal{C}_l|.
\end{equation}
By repeating (\ref{equ6}) and (\ref{equ7}) for $L$ times we then obtain $L$ reflection constellation points as $\tilde{\mathbf{g}}$.
\vspace{-0.3cm}
\subsection{Generalized Gray Coding Method}
After selecting $L$ appropriate reflection constellation points, we propose a generalized Gray coding method to map information bits to constellation by measuring the distance between adjacent constellation points. Specifically, initial point which denoted as $\hat{g}_1$ is randomly selected from $\tilde{\mathbf{g}}$. For current selected reflection constellation $\hat{g}_l$, we calculate the distance between other unselected constellation $\tilde{g}_i$ and $\hat{g}_l$ as $d(\tilde g_i, \hat g_l)$. Thus, we select $\tilde g_{\tilde i}$ having minimum distance to $\hat g_l$ as the next selected reflection constellation $\hat g_{l+1}$, i.e., 
\begin{equation}
\label{gray_coding}
\tilde i=\text{arg}\underset{i}{\text{min}} \, d(\tilde g_i, \hat g_l), \,i=1, 2, ..., L-l.
\end{equation}
Following this, we obtain $L$ ordered reflection constellation points $\hat{\mathbf{g}} $ and map the sequence $\hat{\mathbf{g}}$ to the $b$-bit Gray codes thus ensuring that the Hamming distance of information bits associated with adjacent constellation is $1$.

For the sake of elaboration, we summarize the proposed reflection optimization method in Algorithm 1. It is worth noting that the proposed method is executed only once during a long channel coherence time. Furthermore, by exploiting the statistical channel state information (CSI), one could also optimize the reflection constellation off-line, which is beyond the scope of this letter and reserved for future exploration.

\section{Performance Analysis}

In this section, we will evaluate the error performance of our scheme. Specifically, the average BER of the proposed RIS-aided IM scheme is upper bounded by \cite{refBER}
\begin{equation}
\label{Pe}
P_{\text{e,bit}} \leq \frac{\mathbb{E}_{\mathbf{h}_r^H,\mathbf{G}} \left [ \sum_{\hat l=1}^L \sum_{l=1}^L N(l, \hat l) P(\hat g_l \to \hat g_{\hat{l}}) \right]}{L\text{log}_2L},
\end{equation}
where $N(l, \hat{l})$ denotes the number of error bits between the transmitted symbol $\hat g_l$ and recovered symbol $\hat g_{\hat{l}}$. Based on (\ref{equ4}), the PEP with cascaded reflected channel $\mathbf{Z}$ is given by \cite{PEP}
\begin{equation}
%\label{equ6}
P(\hat g_l \to \hat g_{\hat{l}} | \mathbf{Z}) = P(d_l > d_{\hat{l}} | \mathbf{Z}),
\end{equation}
where $d_l = || y-\sqrt{\rho} \hat g_l ||^2_F $  denotes the decision metric at the receiver for a tentative symbol $\hat g_l$. Furthermore, we have
\begin{equation}
%\label{equ7}
P(d_l > d_{\hat{l}} | \mathbf{Z})
= \int^{\infty}_{\sqrt{\kappa_{l,\hat{l}}}} \frac{1}{\sqrt{2\pi }} e^{- \frac{r^2}{2} } dr = Q(\sqrt{\kappa_{l,\hat{l}}}),
\end{equation}
where $\kappa_{l, \hat{l}} =\frac{\rho}{2\sigma^2} || \hat g_{l} - \hat g_{\hat{l}} || ^2_F$. Due to nonlinear optimization caused by the K-means clustering technique, one can hardly obtain the fixed reflection pattern, which makes it more challenging to get a closed-form expression of (\ref{Pe}). Instead, we next provide an approximate expression numerically.

Specifically, by generating a number of independent and identically distributed (i.i.d.) reflected channels off-line, i.e., $\mathbf{h}_r^H, \mathbf{G},$ and carrying out the same constellation design as discussed in Section III, we could obtain the approximate expression of (\ref{Pe}) as follows
\begin{equation}
\label{equ8}
P_{\text{e,bit}} \le \frac{ \sum^S_{s=1} \sum^{L}_{l=1} \sum^{L}_{\hat{l}=1}
N(l,\hat l)Q(\sqrt{\kappa_{l,\hat{l},s}})
 }{S\, L\, \text{log}_2(L)},
\end{equation}
where $S$ denotes the number of randomizations, and $\kappa_{l, \hat l, s}$ denotes the value of $\kappa_{l,\hat l}$ for the $s$-th independent experiment.

\section{Simulation Results}

In this section, numerical results are provided to verify the improved BER performance of the proposed RIS-aided IM scheme. In our simulations, we consider the phase shift set of $\mathcal{B} = \{ 0, \pi \}$. All simulation results are obtained by averaging $1,000,000$ independent experiments. 

\begin{figure*}[!ht]
\centering
\subfloat[BER performance of the proposed RIS-aided IM scheme ($M=3$).]{\includegraphics[width=5cm]{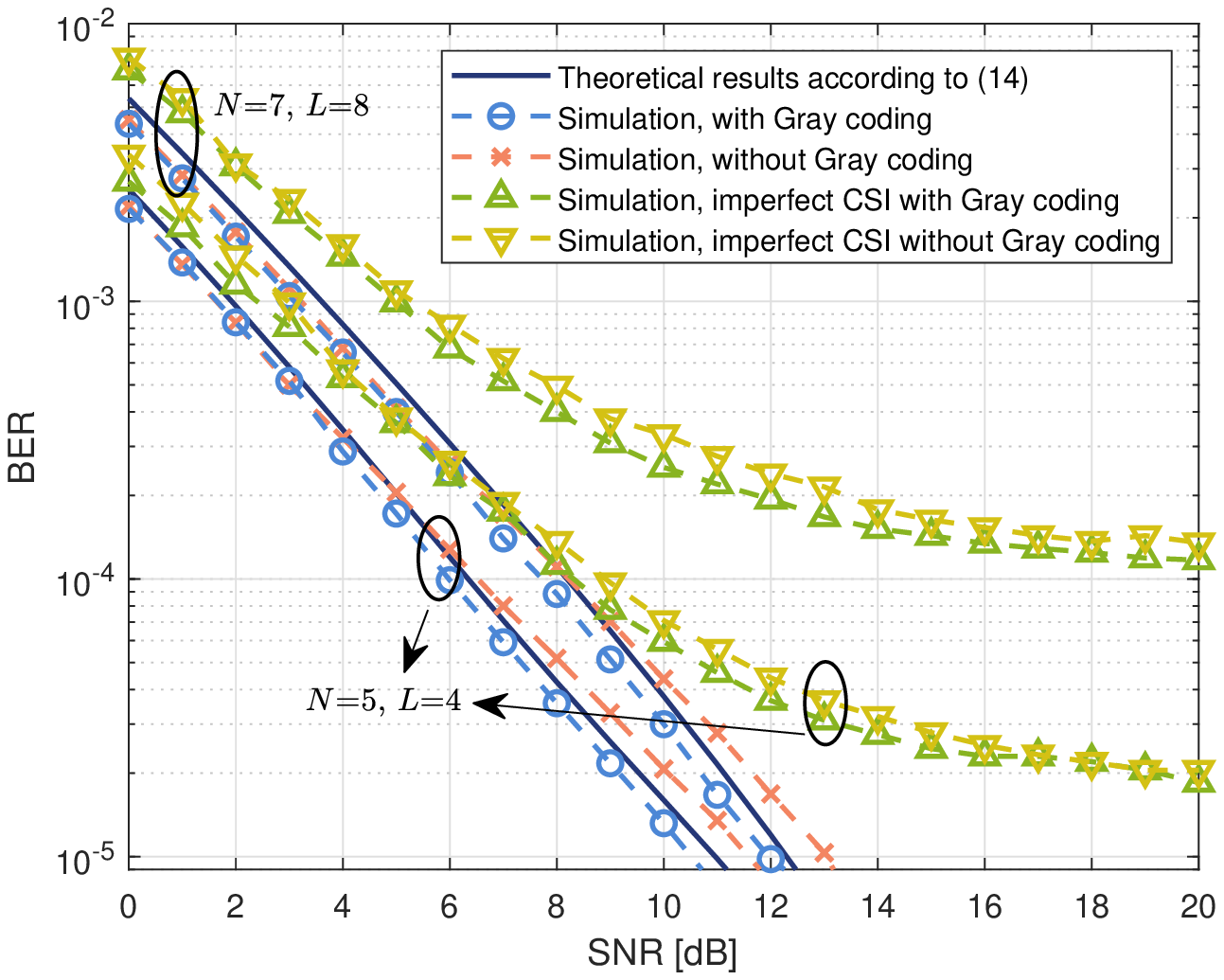}}\qquad
\subfloat[BER performance comparison under different number of RIS elements ($M=3$).]{\includegraphics[width=5cm]{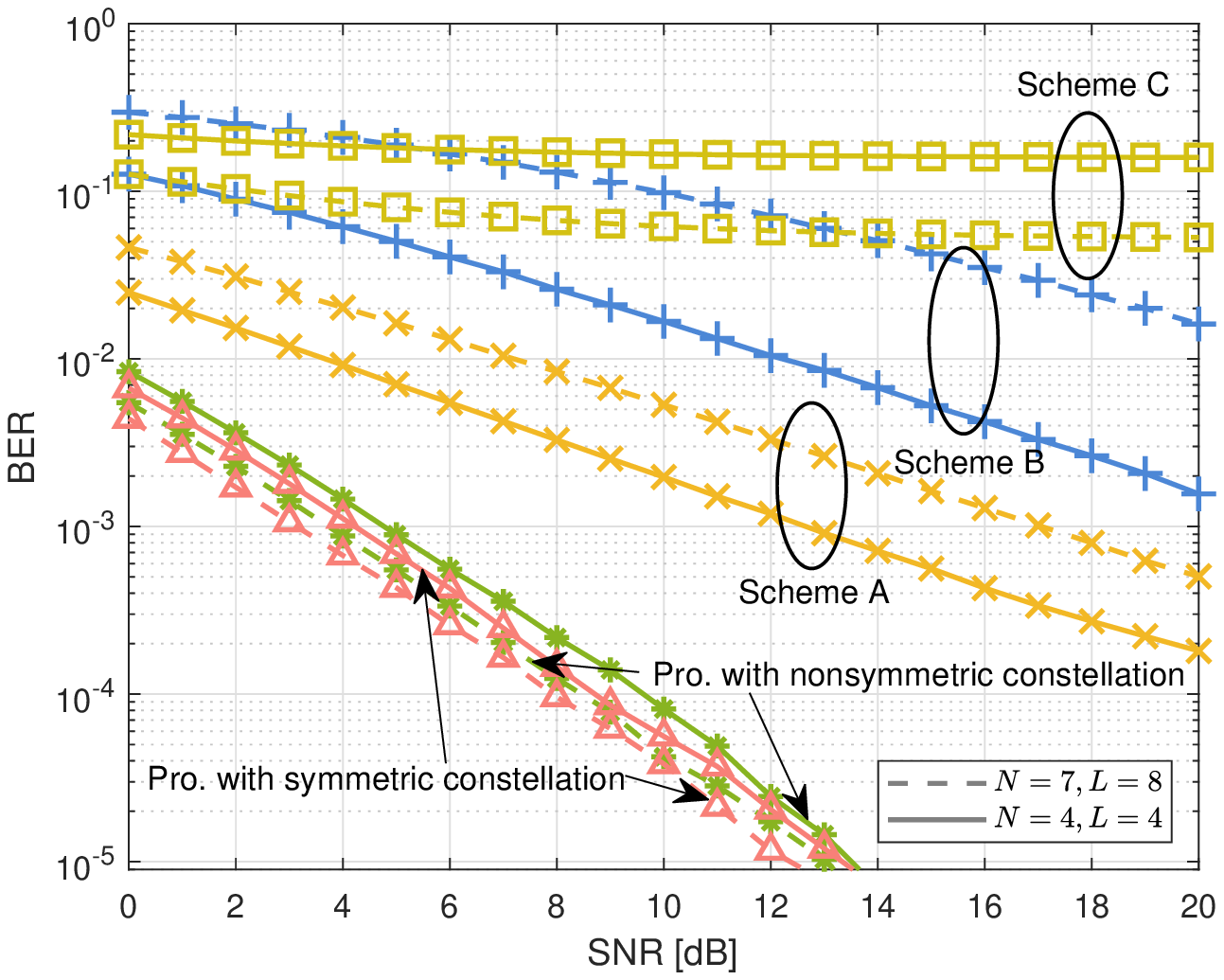}}\qquad
\subfloat[BER performance comparison under different number of transmit antennas ($N=4$, $L=4$).]{\includegraphics[width=5cm]{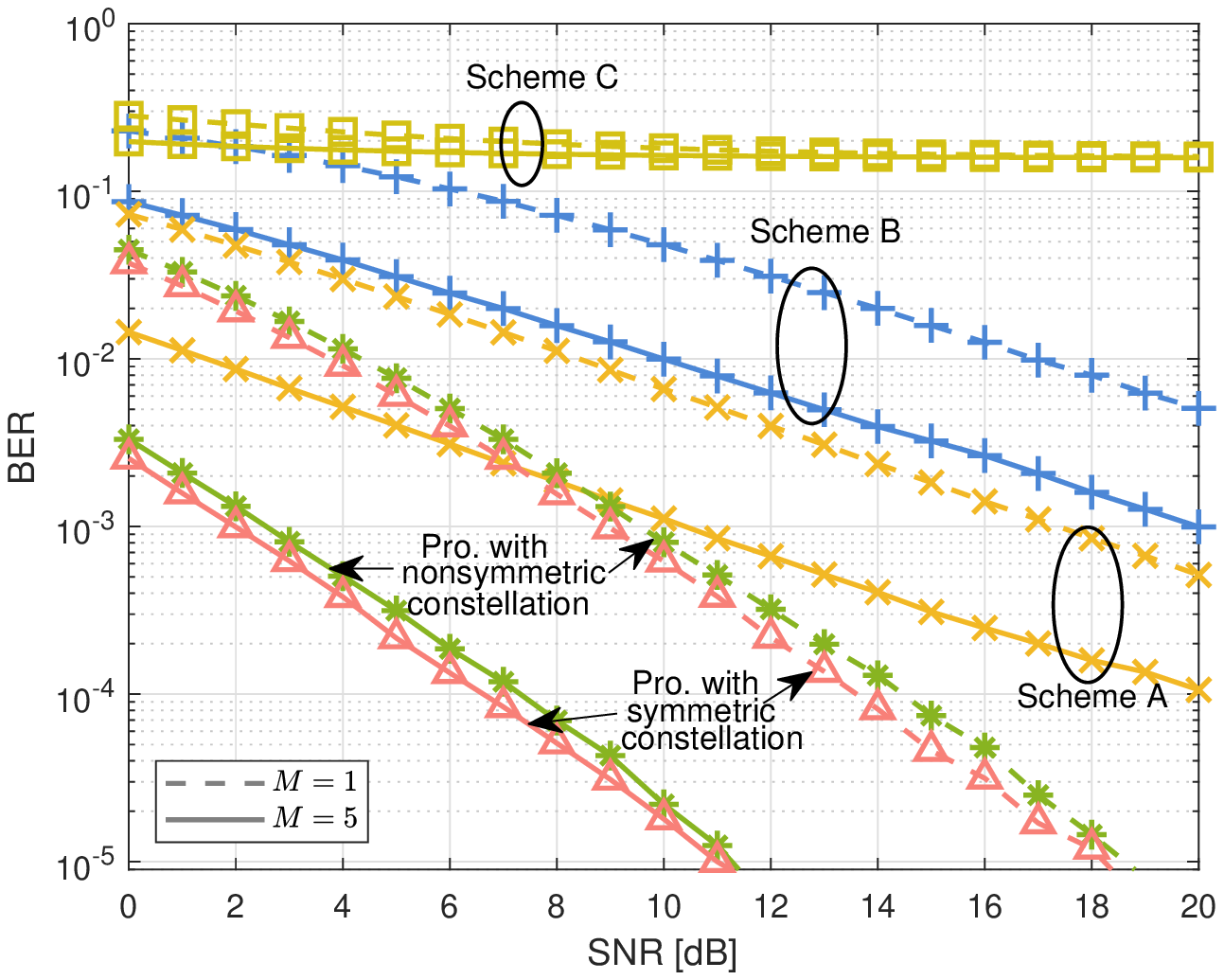}}\vspace{-0.2cm}
\caption{BER performance comparison of various transmission schemes.}\vspace{-0.6cm}
\end{figure*}
For the sake of elaboration, we consider the following three benchmark schemes.

\textbf{\emph{Scheme A}}: Select appropriate reflection constellation based on the sum Euclidean distance maximization philosophy \cite{ref17}.

\textbf{\emph{Scheme B}}: Perform the ON/OFF keying (OOK) scheme by only activating a single RIS element during one time slot \cite{ref10}.

\textbf{\emph{Scheme C}}: Select the reflection constellation randomly.

We first analyze the theoretical performance of the proposed scheme in Fig. 2(a), where the imperfect CSI is also considered. All simulation results are obtained by employing the ML detector at the receiver. The theoretical result is obtained from (14) by taking $S$=1000. For the practical channel estimation errors, we set the average noise power at the BS and the average pilot power to $\sigma_z^2=-50$ dBm and $p_{ul}=-30$ dBm, respectively. Moreover, Fig. 2(a) plots the BER curve of the improved IM scheme without Gray coding. Specifically, we consider the two setups: (i) $N=5$ and $L=4$, (ii) $N=7$ and $L=8$. It is worth noting from Fig. 2(a) that our theoretical analysis serves as a tight upper bound of the proposed scheme under all scenarios considered. Besides, the imperfect CSI would result in an error floor. Nevertheless, the proposed reflection optimization method still gains BER improvements compared to its plain counterpart.

Fig. 2(b) compares the BER performance of the proposed scheme and the three benchmark schemes mentioned above, where we consider two cases: (i) $N=4$ and $L=4$, (ii) $N=7$ and $L=8$. The number of transmit antennas is set to $M=3$. To maintain the same transmission rate, we adopt the binary phase shift keying (BPSK) for the OOK scheme, i.e., \textbf{\emph{Scheme B}}, with $N =7$ RIS elements. It is worth noting from Fig. 2(b) that the randomly selected reflection pattern, i.e., \textbf{\emph{Scheme C}}, resulting the worst BER performance, which hardly gains any improvement with the increase of the SNR. Furthermore, the OOK scheme, i.e., \textbf{\emph{Scheme B}}, performs better than the utterly random scheme by only activating a single element once, which, however, remains performance erosion compared to our scheme. For example, when compared to the best benchmark scheme, i.e., \textbf{\emph{Scheme A}}, the proposed scheme results in a performance gain of more than 12 dB at the BER of $10^{-4}$ and has a significant diversity gain on the slope of the BER curve by selecting the better reflection constellation. In addition, Fig. 2(b) shows the performance improvement of the symmetrical constellation designed in Section III-A over the non-symmetrical design with random phases. Specifically, the symmetrical constellation brings a coding gain of about 1 dB under all setups considered.

In Fig. 2(c), we compare the BER performance of different transmission schemes under the different number of transmit antennas, where we consider two cases: (i) $M=1$, (ii) $M=5$, and adopt $N=4$ and $L=4$ for all setups. Observe from Fig. 2(c) that the BER decreases with the increase of the transmit antennas. Compared to \textbf{\emph{Schemes A}} and \textbf{\emph{B}}, our proposed scheme has a performance gain of 5 dB and 13 dB, respectively, at the BER of $10^{-2}$ benefiting from the optimized reflection constellation. Again, our proposed scheme having symmetric constellation has a 1 dB coding gain compared with the nonsymmetric one for different number of transmit antennas. As the number of transmit antennas increases from $M=1$ to $M=5$, all schemes gain a performance improvement of about 7 dB, while the performance advantage of our proposed scheme remains.

Next, we evaluate the computational complexity of the proposed scheme and benchmark approaches. Specifically, Step A.1 requires calculation of all possible channel gain values, with a complexity of $\mathcal{O}_{A.1}(B^N\log B^N + B^N L)$. In Step A.2, the K-means clustering also has a complexity of $\mathcal{O}_{A.2}(B^N EL)$, where $E$ is the number of iterations. Besides, the complexity of Step A.3 and B are $\mathcal{O}_{A.3}(B^N L-B^N+L)$ and $\mathcal{O}_{B}((L^2-L)/2)$, respectively. Hence, the overall computational complexity of the proposed scheme is given by
\begin{small}
\begin{align}
\mathcal{O}_{pro} &= \mathcal{O}_{A.1} + \mathcal{O}_{A.2} + \mathcal{O}_{A.3} + \mathcal{O}_{B} \nonumber    \\
&\approx \mathcal{O}(B^N \log B^N + B^N EL + B^N L -B^N + L^2).
\end{align}
\end{small}Whereas the computational complexity order of Schemes A, B, and C are $\mathcal{O}_{s.A}(B^N \log B^N + EL^2+L^2)$, $\mathcal{O}_{s.B}(B^N + L \log L)$ and $\mathcal{O}_{s.C}(L)$, respectively. Notably, the proposed scheme provides enhanced BER performance at the expense of increased computational complexity compared to benchmark methods. Further study is needed to balance error performance with computational complexity.

\begin{figure}[!ht]
\centering
\subfloat[Randomly generated constellation.]{\includegraphics[width=3.5cm]{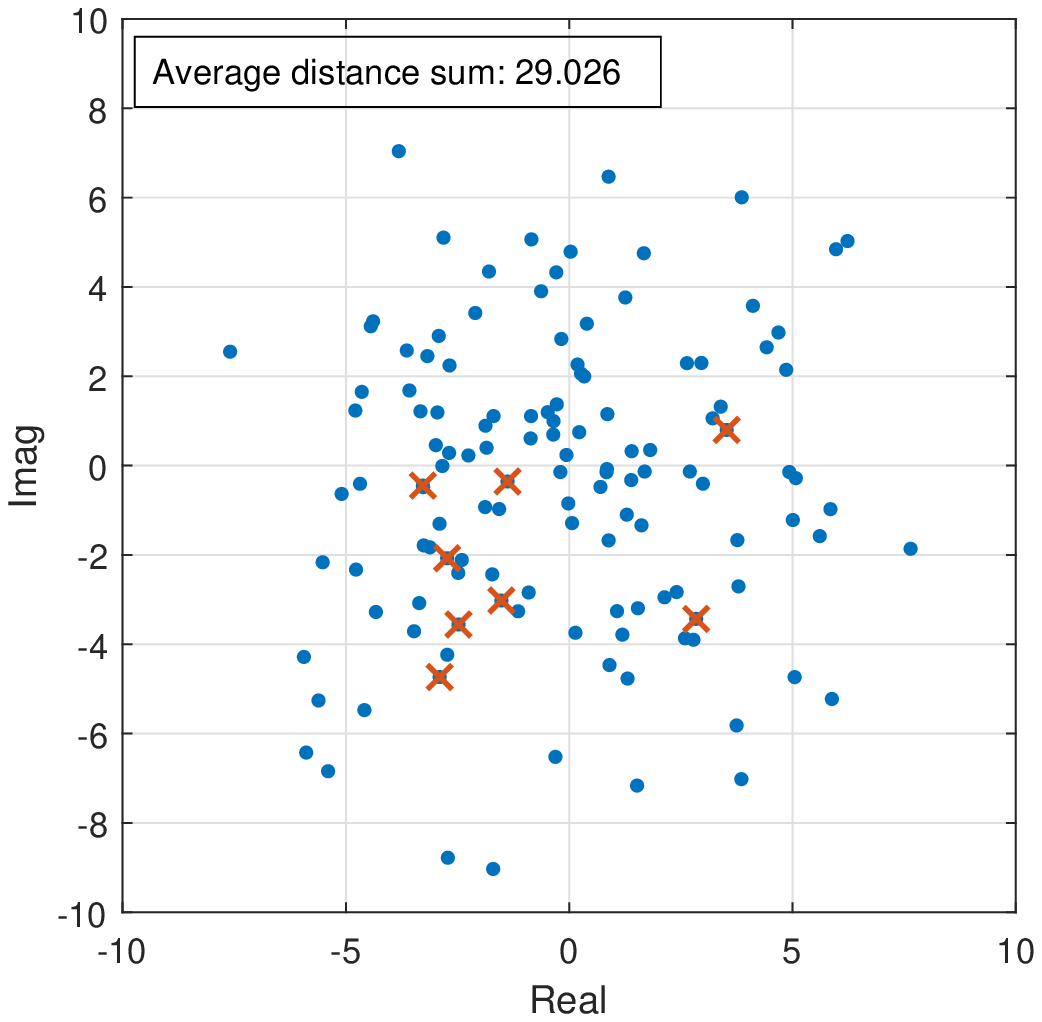}}\qquad
\subfloat[Sum Euclidean distance maximization based constellation.]{\includegraphics[width=3.5cm]{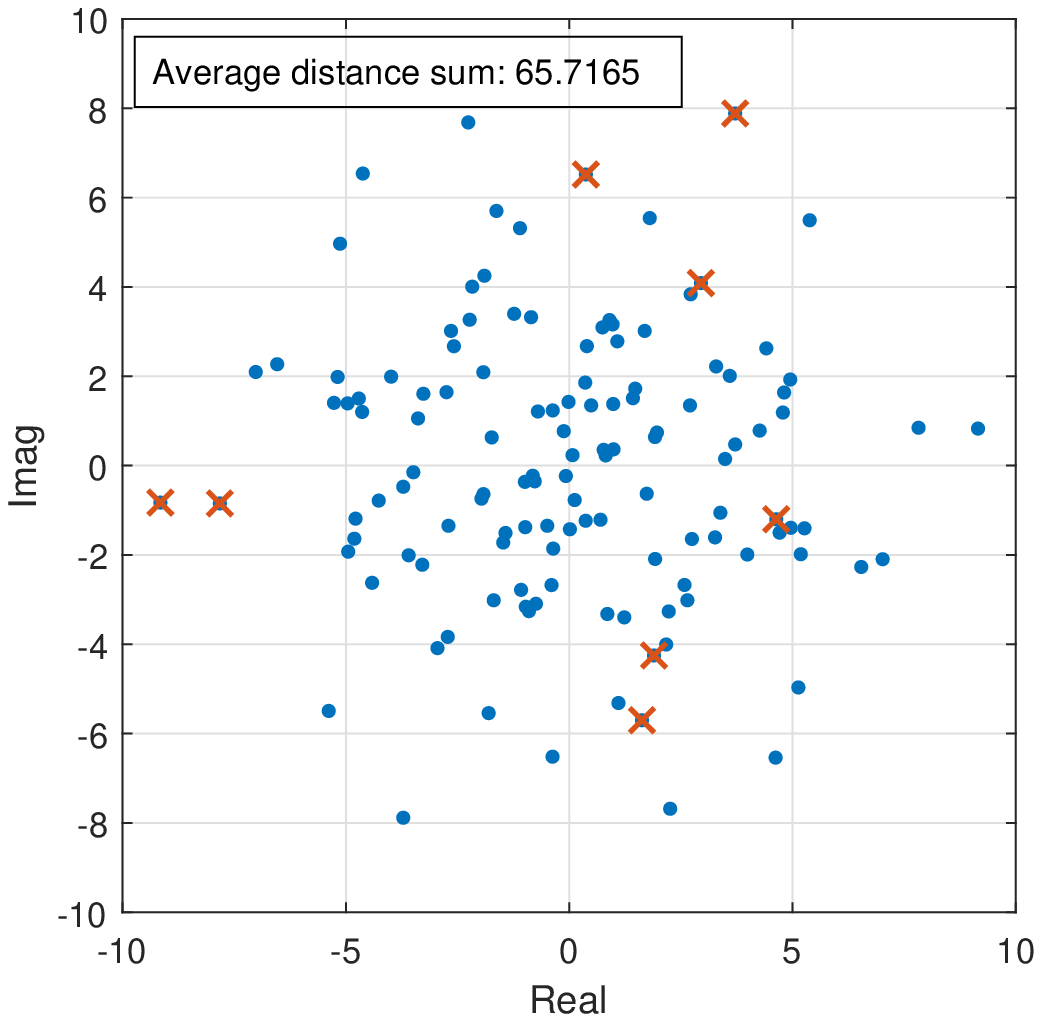}}\\\vspace{-0.3cm}
\subfloat[Proposed K-means based constellation without symmetric constraint.]{\includegraphics[width=3.5cm]{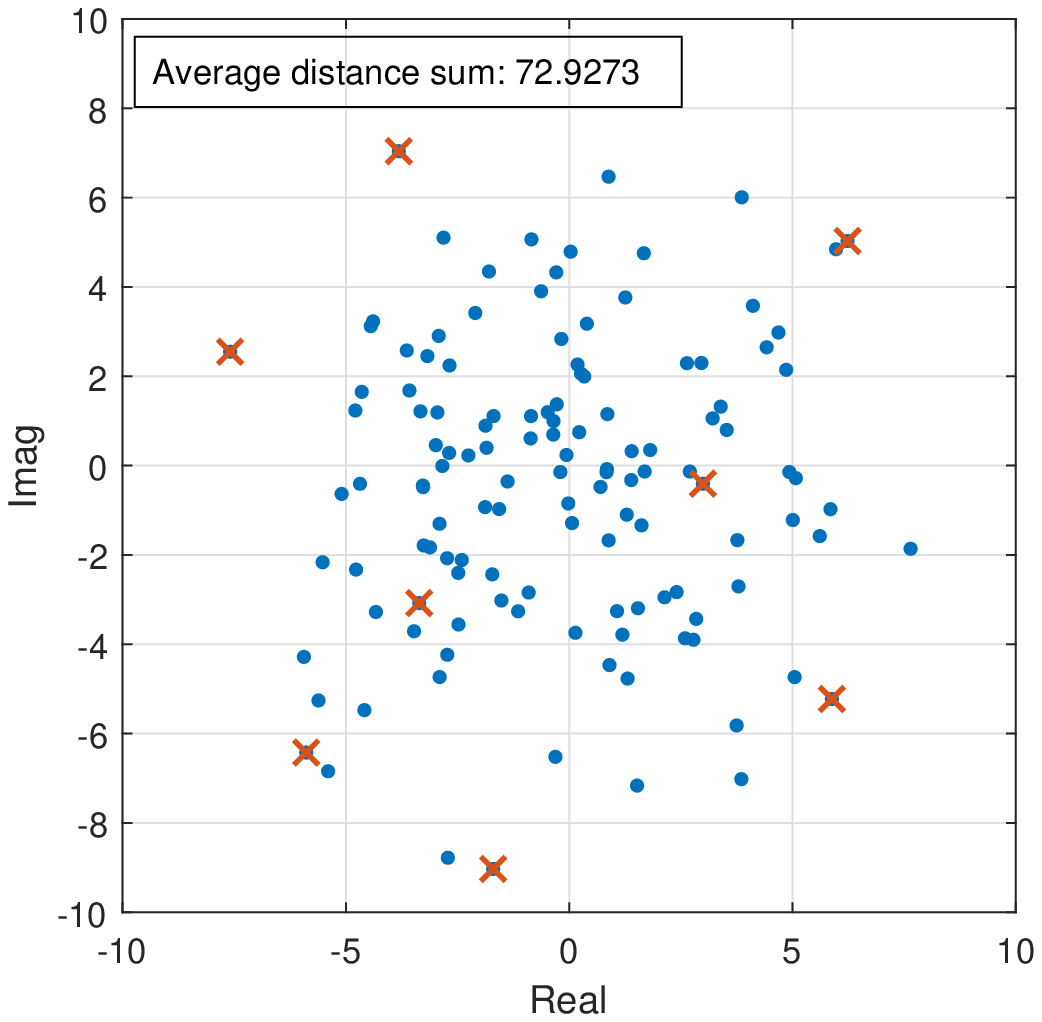}}\qquad
\subfloat[Proposed K-means based constellation with symmetric constraint.]{\includegraphics[width=3.5cm]{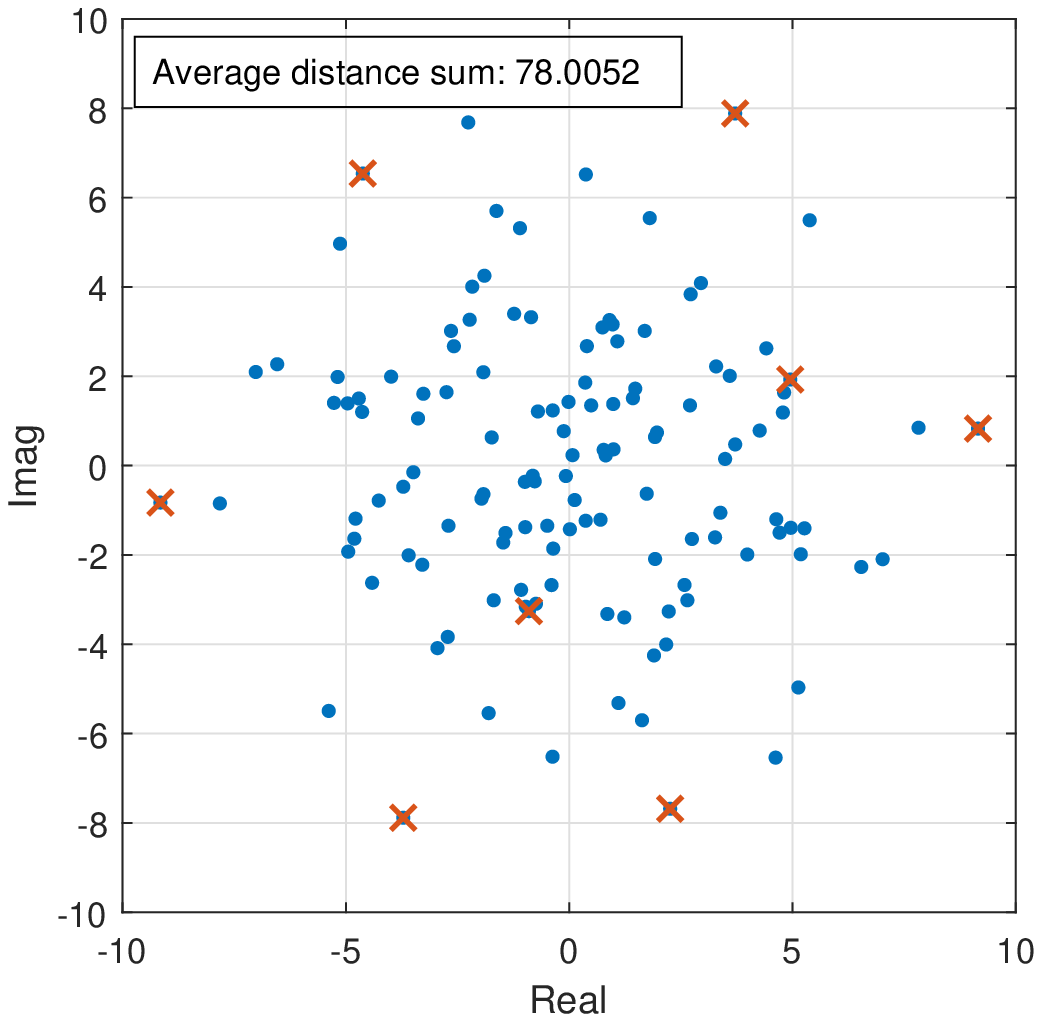}}\vspace{-0.15cm}

\caption{The reflection constellation of different transmission schemes ($R=128$, $L=8$).}\vspace{-0.1cm}
\end{figure}

%\vspace{-5mm}
For the sake of illustration, Fig. 3 shows specific reflection constellation diagrams of different transmission schemes, i.e., $\boldsymbol{\xi}_r^T \mathbf{Z w}_r x_r, \,r=1,2, ..., R,$ where we consider $N=7$, $M=3$, and $L=8$, meaning that 8 patterns are selected from the $R=B^N=2^7=128$ candidate patterns. It can be seen that the randomly selected constellation, i.e., Fig. 3(a), and that produced by applying the sum Euclidean distance maximization policy, i.e., Fig. 3(b), hardly maximize the average distance sum of the reflection constellation, which are about 29 and 65, respectively, thus resulting in BER performance erosion, as verified in Fig. 2. By contrast, the proposed IM scheme could generate better reflection constellation benefiting from the K-means clustering. Specifically, Figs. 3(c) and 3(d) show the constellation of the proposed scheme without and with considering the symmetric constraint in Section III-A, respectively. It is worth noting that by applying the symmetric constraint, the proposed scheme could generate the constellation with a larger average distance sum and achieve better BER performance, i.e., the $1$ dB gain in Fig. 2.
\vspace{-0.2cm}
\section{Conclusion}
In this letter, we proposed a novel IM scheme for the RIS-aided MISO system. In contrast to existing schemes, we employed the K-means clustering technique to optimize the reflection constellation. Furthermore, we proposed a constellation selection approach based on the clustering result and a generalized Gray coding method for mapping information bits to the reflection constellation. Also, we provided an approximate expression to characterize the average BER of the proposed scheme. Numerical results demonstrated the substantially improved BER performance of the proposed scheme compared to existing benchmark schemes. Specifically, our scheme achieves at least $5$ dB performance gain compared to counterparts under all setups considered.
\vspace{-0.2cm}
\bibliographystyle{IEEEtran}
\bibliography{mylib}

\end{document}